# On the tearing instability of highly elongated current sheets

**Grigory Vekstein**

Jodrell Bank Centre for Astrophysics, School of Physics and Astronomy, University of Manchester, Manchester M13 9PL, UK

Email: g.vekstein@manchester.ac.uk

**Abstract.** Rigorous theories of the tearing instability are mathematically quite involving. Therefore, the present note aims to demonstrate how their main results can be reproduced by a simple qualitative analysis of the respective magnetohydrodynamic (MHD) equations.

## 1. Introduction

Plasmoid instability (the tearing mode developing in evolving current sheets) becomes nowadays an integral part in the study of magnetic reconnection [1-5] and magnetic turbulence [6,7] in a highly conducting medium. In this context an important issue is to determine the fastest mode (a mode with the maximal linear instability growth rate), which kick-starts the subsequent process of nonlinear reconnection. It is well-known from the classical Furth-Killeen-Rosenbluth (FKR) theory of the tearing instability [8] that its growth rate increases with the mode wave-length. However, this theory is not applicable to a highly elongated current sheet, which may form in a system with a very large Lundquist number. The respective generalisation of the FKR theory, which is mathematically quite involving, can be found in [9-11]. The present note aims to complement the exact solutions of [9-11] with a simple physical approach, which is based on the order of magnitude estimates of relevant terms in the resistive MHD equations.

The standard procedure in the study of tearing instability involves initial static plasma equilibrium in the magnetic field $\mathbf{B} = B_0 f(x)\hat{\mathbf{y}} + B_T \hat{\mathbf{z}}$. Here $f(x)$, an order of unity odd function of $x$ with a variation length scale $L$, defines the profile of the initial poloidal component with the field reversal at $x=0$. A weak poloidal field perturbation, $\mathbf{b}(x,y,t) = (b_x, b_y)$, can be represented through the flux function $\Psi$ as $\mathbf{b} = (\nabla\Psi \times \hat{\mathbf{z}})$, where for a single Fourier component of the perturbation $\Psi(x,y,t)=\psi(x)\cos ky \exp(\gamma t)$ (here $k$ is the component's wave vector, and $\gamma$ is its growth rate). Such a field perturbation is accompanied with an incompressible plasma flow in the poloidal plane with the velocity $\mathbf{V}(x,y,t) = (\nabla\Phi \times \hat{\mathbf{z}})$, where the respective stream function $\Phi$ takes the form $\Phi(x,y,t)=\phi(x)\sin ky \exp(\gamma t)$. Temporal evolution of perturbations is governed by the resistive MHD equations

$$\frac{\partial \mathbf{B}}{\partial t} = \nabla \times (\mathbf{V} \times \mathbf{B}) + \eta \nabla^2 \mathbf{B}, \quad \text{and} \quad \rho \frac{d\mathbf{V}}{dt} = \frac{1}{4\pi}(\nabla \times \mathbf{B}) \times \mathbf{B} - \nabla P$$

By taking *curl* of the of the equation of motion, and re-writing these equations in the linear approximation with respect to weak perturbations, one arrives to the following set of equations for the functions $\psi(x)$ and $\phi(x)$:

$$\gamma \psi(x) = k B_0 f(x) \phi(x) + \eta(\psi'' - k^2 \psi), \tag{1}$$

$$\gamma(\phi'' - k^2 \phi) = -\frac{k B_0}{4\pi \rho}[f(\psi'' - k^2 \psi) - f'' \psi] \tag{2}$$

The very development of the tearing instability, which involves reconnection of magnetic field lines, is not possible without a finite plasma resistivity. Therefore, when the resistivity is weak, i.e. the respective Lundquist number $S \equiv L V_A / \eta \gg 1$, ($V_A = B_0 / \sqrt{4\pi \rho}$ is the characteristic Alfven velocity), the tearing mode is slow in the plasma inertial timescale $\tau_A = L / V_A$: $\gamma \tau_A \ll 1$. Thus, in the main body of the system its evolution is quasi static, hence, by ignoring the inertial term on the l.h.s. of Eq.(2), one arrives to the following equation for $\psi(x)$:

$$\psi'' - (k^2 + f'' / f)\psi = 0 \tag{3}$$

In this region a weak plasma resistivity is also insignificant, so it follows then from Eq.(1) that the stream function $\phi(x) = \gamma \psi(x) / k B_0 f(x)$. Note that while being slow in the inertial timescale, the tearing instability is fast in the global resistive timescale $\tau_\eta = L^2 / \eta = \tau_A S \gg \tau_A$, hence $\tau_\eta^{-1} \ll \gamma \ll \tau_A^{-1}$.

It turns out, however, that under the condition that perturbations vanish at external boundaries, in general case, there is no regular non-trivial solution of Eq.(3). Therefore, tearing mode corresponds to a singular solution of this equation, with a discontinuity of $\psi'$, i.e. of $b_y$, which should be located at $x = 0$, the reversal point of the initial field. This discontinuity of $b_y$, i.e. the emerging current sheet, is characterised by the parameter

$$\Delta' = \frac{\psi'(0+\varepsilon) - \psi'(0-\varepsilon)}{\psi(0)}, \tag{4}$$

which is at the heart of the tearing instability. Indeed, it is well-known (see, e.g., [12]), that the amount of magnetic energy released by the tearing-like perturbation, $\Delta W_M$, is equal to

$$\Delta W_M = \Delta'[\psi(0)]^2 / 8\pi. \tag{5}$$

Therefore, growth of the tearing-like perturbation is energetically favourable, i.e. the system is tearing unstable, if $\Delta' > 0$. A particular value of this parameter depends on both the profile of the initial magnetic field [given by the function $f(x)$], and the perturbation wave-vector $k$. For example, one can easily verify that in the case of a linear field reversal, when $f(x) = x/L$, $\Delta'$ is negative whatever the wave-vector, i.e. such a field is always tearing stable. On the other hand, for the standard Harris profile with $f(x) = \tanh(x/L)$, one gets [see, e.g., 12] $\Delta' = 2(1-k^2L^2)/kL^2$. Thus, such a field is unstable for long wave-length perturbations with $kL < 1$, but remains stable for short wave-length modes.

Although the energetics of the tearing instability is entirely defined by a proper solution of Eq.(3) (the so-called external solution), the instability growth rate, $\gamma$, is determined by the plasma dynamics inside the current sheet, a width of which becomes non-zero when a weak but finite plasma resistivity (and plasma inertia) are accounted for. The respective, internal, solution of Eqs.(1-2) is discussed in the next Section.

## 2. The internal solution and the instability growth rate.

Under a large value of the Lundquist number, $S \gg 1$, a width of the formed current sheet, $\Delta x$, is small: $\Delta x \ll L$ (see below), which allows to simplify Eqs.(1-2) in the following way. Firstly, in this case the profile function $f(x)$ can be approximated as $f(x) \approx x/L$ (this is, actually, a rigorous definition of the scale-length $L$ introduced above). Secondly, within such a narrow current sheet $\psi'' \gg k^2\psi$, $\phi'' \gg k^2\phi$ (it is clear from what follows that only $k \leq L^{-1}$ are of interest), hence, instead of Eqs.(1-2), one now gets

$$\gamma\psi = \frac{kB_0 x}{L}\phi + \eta\psi'', \quad \gamma\phi'' = -\frac{kB_o x}{4\pi\rho L}\psi'' \qquad (6)$$

In dimensionless variables defined by the following re-scaling:

$$x \to x/L, k \to kL, \psi \to B_0 L, \phi \to V_A L, \qquad (7)$$

Eqs.(6) take the standard tearing theory form

$$(\gamma\tau_A)\psi = kx\phi + S^{-1}\psi'', \qquad (8a)$$

$$(\gamma\tau_A)\phi'' = -kx\psi'' \qquad (8b)$$

These equations yield two different regimes of the tearing instability. The first one is the well-known FKR regime [8], which rely on the so-called "constant-psi" approximation. The second regime corresponds to long-wave modes with a very large value of the tearing parameter $\Delta'$ (which in dimensionless variables is measured in units of $L^{-1}$), when the "constant-psi" approach becomes not applicable. The respective solution was originally obtained in [9], and, hence, termed the Coppi solution. An elegant exact general solution that incorporates both these regimes is presented in a recent publication [11]. Although the emphasis of the present note is on the Coppi regime, it is helpful to start with reproducing the FKR results by exploring a simple qualitative analysis of Eqs. (8).

## A. The FKR ("constant-psi") regime

In this case one can, without loss of generality, put $\psi = 1$, so the matching condition of the internal and external solutions then reads $\Delta' = \int \psi'' dx$. For the current sheet of width $\Delta x$ the latter integral can be estimated as $\psi'' \cdot \Delta x$, which yields

$$\psi'' \sim \Delta'(\Delta x)^{-1} \tag{9}$$

Further on, by using the equation of motion (8b), one can estimate the stream function in the internal solution. Thus, $\phi'' \sim \phi(\Delta x)^{-2}$, while, according to (9), $x\psi'' \sim \Delta x \cdot \psi'' \sim \Delta'$. Hence, it follows from (8b) that

$$\phi \sim k\Delta'(\gamma \tau_A)^{-1}(\Delta x)^{-2} \tag{10}$$

Consider now the magnetic induction equation (8a). The ongoing magnetic reconnection, the pace of which is defined by the l.h.s. of this equation, is supported both by the plasma resistivity (the second term on the r.h.s.) and the advection of magnetic field into the current sheet (the first term on the r.h.s.). Therefore, all three terms of Eq.(8a) should be of the same order of magnitude. Thus, by comparing the last two with the help of (9) and (10), one gets: $k^2\Delta'(\Delta x)^3(\gamma\tau_A)^{-1} \sim S^{-1}\Delta'(\Delta x)^{-1}$ hence,

$$\Delta x \sim S^{-1/4}(\gamma\tau_A)^{1/4} k^{-1/2} \tag{11}$$

Finally, the requirement that the other two terms are also of the same order of magnitude, namely

$$(\gamma\tau_A)\psi = (\gamma\tau_A) \sim S^{-1}\psi'' \sim S^{-1}\Delta'(\Delta x)^{-1} \tag{12}$$

yields, together with (11), the well-known results of the FKR theory [8]:

$$(\gamma\tau_A) \sim S^{-3/5}(\Delta')^{4/5}k^{2/5} \;,\; (\Delta x) \sim S^{-2/5}(\Delta')^{1/5}k^{-2/5} \tag{13}$$

Consider now what restrictions apply to this solution by the imposed "constant-psi" assumption. Clearly, the variation of the flux function $\psi(x)$ across the current sheet can be estimated as $\Delta\psi \sim \psi'' \cdot (\Delta x)^2$, which should remain small: $\Delta\psi \ll 1$. With the help of (9) and (13) this requirement takes the form:

$$\Delta'(k) < S^{1/3}k^{1/3} \qquad (14)$$

Since the tearing parameter $\Delta'(k)$ is usually increasing with the wave-vector $k$ getting smaller, it follows from (14) that the "constant-psi" FKR solution holds for $k > k_*$, with the latter defined by the condition

$$\Delta'(k_*) \sim S^{1/3}k_*^{1/3} \qquad (15)$$

For example, in the case of the Harris equilibrium, for which $\Delta'(k) \sim k^{-1}$, Eq.(15) yields the well-known $k_* \sim S^{-1/4}$. Likewise, for the sine-like profile considered in [11], when $\Delta'(k) \sim k^{-2}$, one gets $k_* \sim S^{-1/7}$. More generally, if the tearing parameter scales as $\Delta'(k) \sim k^{-\alpha}$ with $\alpha > 0$, it follows then from (15) that

$$k_* \sim S^{-1/(3\alpha+1)} \qquad (16)$$

Finally, note that according to (13) and (15)

$$\gamma(k_*)\tau_A \sim S^{-1/3}k_*^{2/3}, \quad \Delta x(k_*) \sim S^{-1/3}k_*^{-1/3} \qquad (17)$$

By the very meaning of the separation of the external and internal solutions, the width of the current sheet should be small: $\Delta x \ll 1$. According to (18), it requires $k_* > S^{-1}$, which is indeed the case as seen from (16).

## B. The Coppi regime (non "constant-psi" solution)

For a long-wave mode with $k < k_*$ the inequality opposite to (14) holds:

$$\Delta'(k) > S^{1/3}k^{1/3}, \qquad (18)$$

which makes the "constant-psi" approximation non-applicable. Therefore, in this case one has to distinguish between the $\psi_e(0)$, which is the limit of the external solution at $x \to 0$, and $\psi_i(0)$, which is defined by the internal solution. Note that the former determines the free magnetic energy associated with the tearing perturbation [see Eq.(5)], while the latter defines the amount of reconnected magnetic flux and, hence, the size of magnetic islands formed inside the current sheet (see, e.g.,[12]).

In what follows we put, as before, $\psi_e(0) = 1$, and denote $\psi_i(0)$ as $\psi_i$. Therefore, the above-derived relations (9-11) remain unchanged, while in Eq.(12) the former $\psi$ should be now replaced with $\psi_i$ - the actual reconnected magnetic flux. Hence, instead of (12), one gets

$$(\gamma\tau_A)\psi_i \sim S^{-1}\psi'' \sim S^{-1}\Delta'(\Delta x)^{-1} \qquad (19)$$

Furthermore, since inside the current sheet the flux function is now not a constant, its total variation across the current sheet, which is equal to $\Delta\psi = (\psi_i - \psi_e) = (\psi_i - 1)$ should be accounted for as $(\psi_i - 1) \sim \psi'' \cdot (\Delta x)^2 \sim \Delta' \cdot (\Delta x)$ [see Eq.(9)]. Since, as shown below, in this regime $\psi_i \gg 1$, it simply yields $\psi_i \sim \Delta' \cdot (\Delta x)$, and, with the help of (11) and (19), one finally gets

$$\Delta x \sim S^{-1/3} k^{-1/3} \;,\; (\gamma \tau_A) \sim S^{-1/3} k^{2/3} \tag{20}$$

Remarkably, in this regime of the tearing mode its growth rate and the width of reconnective current sheet do not depend on the tearing parameter $\Delta'$ (provided, of course, that the latter is large enough so that inequality (18) is satisfied). Note also that this inequality also ensures that $\psi_i \sim \Delta' \cdot \Delta x$ is large: $\psi_i \gg 1$. Furthermore, as seen from (17) and (20), the two regimes, FKR and Coppi, match each other at $k \sim k_*$ [with the latter defined in Eq.(15)]. Hence, the fastest growing tearing mode is the one with $k \sim k_*$.

Finally, two remarks concerning the range of validity of the Coppi regime. Firstly, the condition $\Delta x \ll 1$ should be met, which, according to (20), requires

$$k > k_{min} \sim S^{-1} \tag{21}$$

The second one is about the quasi static assumption imposed on the external solution. The point is that the characteristic spatial scale for a mode with a wave-number $k \ll 1$ is equal to $\lambda \sim L \cdot k^{-1} \gg L$. Therefore, the respective inertial time-scale $\tau_A^{(\lambda)} \sim \tau_A k^{-1} \gg \tau_A$, which makes the quasi static criterion more restrictive, namely $\gamma(k) \cdot \tau_A^{(\lambda)} \sim \gamma(k) \tau_A k^{-1} < 1$. Nevertheless, as seen from (20), it remains satisfied under the condition (21).